\documentclass[11pt]{article}
\usepackage{graphicx} 
\usepackage[utf8]{inputenc}
\usepackage{amsfonts,url,epsfig,float,xcolor}
\usepackage{amsmath}
\usepackage{geometry} 
\usepackage{sectsty} 
\usepackage[normalem]{ulem}
\usepackage{gensymb}
\usepackage{subcaption}

\usepackage[style=authoryear-comp]{biblatex}
\sectionfont{\sffamily\bfseries\upshape\large}
\subsectionfont{\sffamily\bfseries\upshape\normalsize} 
\subsubsectionfont{\sffamily\mdseries\upshape\normalsize}
\makeatletter
\renewcommand\@seccntformat[1]{\csname the#1\endcsname.\quad}
\makeatother

\makeatletter
\def\@maketitle{%
  \begin{center}%
  \let \footnote \thanks
    {\large \@title \par}%
    {\normalsize
      \begin{tabular}[t]{c}%
        \@author
      \end{tabular}\par}%
    {\small \@date}%
  \end{center}%
}
\makeatother
 \usepackage{lineno}
\title{\bf Russian roulette: The need for stochastic potential outcomes when utilities depend on counterfactuals \vspace{.1in}}

\author{Andrew Gelman\footnote{Departments of Statistics and Political Science, Columbia University, New York, ag389@columbia.edu} \and Jonas M. Mikhaeil\footnote{Department of Statistics, Columbia University, New York, j.mikhaeil@columbia.edu}\vspace{.1in}}

\date{27 Jun 2025\vspace{-.1in}}

\begin{document}

\maketitle

\begin{abstract}
It has been proposed in medical decision analysis to express the ``first do no harm'' principle as an asymmetric utility function in which the loss from killing a patient would count more than the gain from saving a life. Such a utility depends on unrealized potential outcomes, and we show how this yields a paradoxical decision recommendation in a simple hypothetical example involving games of Russian roulette.  The problem is resolved if we abandon the stable unit treatment value assumption (SUTVA) and allow the potential outcomes to be random variables. This leads us to conclude that, if you are interested in this sort of asymmetric utility function, you need to move to the stochastic potential outcome framework.  We discuss the implications of the choice of parameterization in this setting.
\end{abstract}

\section{Introduction}
\subsection{Utility functions for potential outcomes in causal inference}


Causal inference has traditionally focused on identification of average effects.  In recent decades, there has been increasing interest in considering different forms of the average causal effect \parencite{imbens1994identification}, flexible models for individual effects \parencite{hill2011bayesian,wager2018estimation}, and understanding how effects vary \parencite{ding2019decomposing}.

Individual causal effects can never be precisely estimated (without multiple measurements on the same units and strong assumptions about crossover effects), but they can be defined theoretically as latent variables using the potential-outcomes framework \parencite{neyman1923application,rubin1974estimating}.  In the simplest case of binary treatment $z$ and binary outcome $y$, the potential outcomes for unit $i$ are $(y_i^0,y_i^1)$, and these partition the population into four classes or principal strata \parencite{frangakis2002principal}, which can be modeled as latent outcomes \parencite{page2015principal}.

\textcite{benmichael2024policy} and \textcite{christy2024starting} have argued that, once effects are defined (even if only implicitly) at the individual level, it can make sense to consider asymmetric losses using some formalization of the ``first do no harm'' principle.  Analyses using an asymmetric loss function have potential practical implications, because the average utility is no longer the same as the average causal effect. Utility functions over potential outcomes have also been considered by \textcite{cui2024policy} and \textcite{guggenberger2024minimax}. 

\subsection{Relation to classical utility theory}

There can be reasonable policy or ethical reasons to prefer low-risk treatments, but we have to be careful, as it is well known that any utility function that depends on unrealized outcomes can lead to incoherent decision recommendations; see, e.g., \parencite{bell1982regret}.

For a simple example of an incoherent decision rule when moving beyond the classical outcome-based utility framework of \textcite{vonneumann1944theory}, consider a decision between two options:  A, a lottery which yields \$100 with probability 0.6 and \$0 with probability 0.4; and B, which with probability 0.5 yields \$100 and with probability 0.5 returns a lottery which yields \$100 with probability 0.2 and \$0 with probability 0.8.  These two lotteries are identical in the distribution of outcomes and thus would receive identical utilities in the classical Neumann-Morgenstern framework.  But now consider a utility function in which there is a negative value to uncertainty, which is evaluated at each branch of the tree.  For example, suppose the utility of \$100 is set to 1, the utility of \$0 is set to 0, and the existence of uncertainty reduces the utility of any lottery by 10\%.  Then option A has utility $0.9 \,(0.6\cdot 1 + 0.4\cdot 0)= 0.54$, while option B has utility $0.9\, (0.5\cdot 1 + 0.5\,(0.9\,(0.2\cdot 1 + 0.8\cdot 0))) = 0.9\,(0.5 + 0.09) = 0.531$.  Option B here has a lower utility because the utility function pays twice for the uncertainty.  This is not to say that such a utility function is ``wrong'' (or, for that matter, that it is normatively or descriptively right); it just illustrates how it violates the axioms of classical (Neumann-Morgenstern) decision theory.

This example is {\em not} just a simple nonlinear utility function for gains and losses (which has its own coherence problems; see section 5 of \parencite{gelman1998demonstrations}); rather, it violates classical decision theory by depending on unrealized outcomes.

When a decision rule is incoherent, it should be possible to construct a ``money pump'' or some other scenario that yields inappropriate or nonsensical decision recommendations.  Again, that does not mean the rule is never appropriate.  Working through simple examples can help us understand its domain of applicability.

\subsection{Expressing the ``first do no harm'' principle as an asymmetric utility function}

Suppose we want to decide between two possible treatments, where $z=0$ is the ``control'' or status quo and $z=1$ is a proposed new treatment. We follow  \textcite[Section 3.1]{christy2024starting} in using a decision rule based on potential outcomes $(y^0,y^1)$, where $y^0$ is the outcome under $z=0$ and $y^1$ under $z=1$. Specifically, we assume the following relative utility function:
\begin{equation}\label{utility_discrete}
U(z=1\mbox{ relative to } z=0\,|\,y^0,y^1) = \left\{
  \begin{array}{rl} 0 & \mbox{if $U(y^1)=U(y^0)$}\\
 0.5(U(y^1) - U(y^0)) & \mbox{if $U(y^1) > U(y^0)$}\\
 - (U(y^1) -U( y^0)) & \mbox{if $U(y^0) > U(y^1)$,}
\end{array}\right.
\end{equation}

\noindent
so that the disutility of a loss is twice as large as the utility of a gain. We introduce the notation ``relative to'' to highlight the asymmetry of this utility function designed to favor the status quo.

Consider the case in which $y=1$ is survival and $y=0$ is death.  Then the four principal strata correspond to $(y^0,y^1)=(1,1)$ (patient survives under either treatment), $(0,0)$ (patient dies under either treatment), $(0,1)$ (treatment saves the patient's life), and $(1,0)$ (treatment kills the person).  In the first two principal strata, the treatment effect is zero.  If the negative of the utility in the fourth stratum is taken to be greater than the positive utility of the third stratum, then the resulting decision rule will ``prioritize safety over efficacy'' \parencite{christy2024starting}.  Without loss of generality, we can write $U(0)=0$ and $U(1)=1$, and equation (\ref{utility_discrete}) becomes,
\begin{equation}\label{utility_discrete_2}
U(z=1\mbox{ relative to } z=0\,|\,y^0,y^1) = \left\{\begin{array}{rl} 0 & \mbox{if you are sure to live} \\
                   0 & \mbox{if you are sure to die} \\
                   +0.5 & \mbox{if the treatment would save your life} \\
                   -1 & \mbox{if the treatment were to kill you.}
                        \end{array}\right.
\end{equation}

\section{Russian roulette}

In the case of the ``do no harm'' principle, we can demonstrate the incoherence of the above asymmetric utility function using a simple example of two independent lotteries:
\begin{itemize}
\item $R_{1/6}$:  You live with probability 5/6 and die with probability 1/6,
\item $R_{1/7}$:  You live with probability 6/7 and die with probability 1/7.
\end{itemize}
It is a clear improvement to switch from the status quo  ``control'' of $R_{1/6}$ to the ``treatment,'' $R_{1/7}$.  But, as we shall see, this is not necessarily recommended under a regret-based decision rule.

We assume the outcomes in $R_{1/6}$ and $R_{1/7}$ are independent, so we can easily work out the probabilities of the four potential strata:  $(y^0,y^1)$ takes on the value $(1,1)$ with probability 30/42, it takes on the value $(0,0)$ with probability 1/42, it is $(1,0)$ with probability 5/42, and it is $(0,1)$ with probability 6/42. Under expression (\ref{utility_discrete_2}), the expected relative utility of $R_{1/7}$ compared to $R_{1/6}$ is thus $0.5(6/42) - 5/42= -1/21$.  Here the potential outcomes are fixed but unknown quantities, and the probabilities correspond to uncertainty about the states of the guns or, equivalently, as averages over a hypothetical superpopulation of plays.

This result---that we should stick with $R_{1/6}$---makes no sense.  Obviously we would prefer our chance of dying to decrease from 1/6 to 1/7.  But, under the asymmetric utility function, you lose more from the 5/42 chance of the new treatment ``killing you'' than you gain from the 6/42 chance of it ``saving your life.''

We can also work out the above algebra under an alternative formulation. Let $\phi$ be a random variable that equals 1 if the chamber is loaded with a bullet when the player draws the trigger and 0 otherwise. We then have the deterministic potential outcomes $y(\phi=0) = 1$ (alive) and $y(\phi=1) = 0$ (dead). In this parameterization, all randomness is induced by the distribution over $\phi$ induced by the treatment $z$. Different lotteries $R_{p}$ correspond to different distributions of the intermediate treatment, $\phi$. In this situation, $(y^0,y^1) = (y(\phi),y(\phi^\prime)) $ with independent random variables $\phi\sim\mbox{Bernoulli}(1/6)$ and $\phi^\prime\sim\mbox{Bernoulli}(1/7)$.
Expression (\ref{utility_discrete}) then becomes,
\begin{align} \nonumber
    \mbox{E}_{\phi,\phi^\prime} \left( U(R_{1/7}\mbox{ relative to } R_{1/6}\,|\,y^0,y^1)\right ) 
  \,  &= \, \mbox{E}_{\phi,\phi^\prime} \left(0.5 \cdot 1_{y^1 > y^0} - 1_{y^0 > y^1} \right) \\\nonumber
    \,  &= \,  0.5 \cdot\mbox{Pr}(\phi > \phi^\prime) - \mbox{Pr}(\phi^\prime > \phi ) \\\nonumber
    &= 0.5(1/7) - 5/42= -1/21,
\end{align}

\noindent
and again the asymmetric decision rule yields an unacceptable conclusion.

\section{Moving from deterministic to stochastic potential outcomes}

There are two natural ways to resolve this problem and avoid the nonsensical decision recommendation.  The first approach would be to move to a classical utility function that depends only on outcomes, in which case one can just evaluate $U(R_{1/6})$ and $U(R_{1/7})$, and there is no incoherence.  This solution is too easy, though, in that it eliminates the ``do no harm'' principle that motivated the idea in the first place.

A second approach preserves the asymmetric utility but changes the nature of the potential outcomes, replacing the {\em deterministic} potential outcomes of \textcite{neyman1923application} and \parencite{rubin1974estimating} with {\em stochastic} potential outcomes as in \parencite{greenland1987effect} and \parencite{vanderweele2012stochastic}.  

To adapt (\ref{utility_discrete}) to allow stochastic potential outcomes $y^z$ (where $z=0$ or 1, corresponding to the $R_{1/6}$ or $R_{1/7}$, respectively), the utility of a potential outcome now needs to take its variability into account. In the Russian roulette example, we replace each binary outcome $y^z$ with a binary random variable whose expectation is the probability of survival under treatment $z$.  The Russian roulette treatment thus includes the spinning of the cylinder as well as the pulling of the trigger, and the assumption is that the outcome of the spin is random and is not predictable by any characteristics of the individual.
It's natural to average over this level of variability when evaluating the utilities $U(y^z)=E(y^z)$, that is, to define the utility as the probability of survival.  What is relevant here is that the utilities of $y^0$ and $y^1$ are evaluated individually, and then the asymmetric utility for the decision is applied to these separate utilities.  This follows the same principles as in the previous section, with the only change being that the potential outcomes $y^0$ and $y^1$ are random variables.

When evaluating $R_{1/7}$ versus the status quo of $R_{1/6}$, the utilities are $U(y^1)=6/7$ and $U(y^0)=1/6$ for everyone in the population, so the decision will certainly {\em increase} survival probability.  The expected relative utility is $0.5\,(6/7-5/6) = 1/84$, which is positive, as it should be.

When switching from deterministic to stochastic potential outcomes the order between evaluating utilities and integration over uncertainty changes. In the deterministic case, we first evaluate the utilities and then integrate over the possible treatment assignments under the policies $R_p$. When using stochastic potential outcomes, the utilities are evaluated on random variables, collapsing the uncertainty in a unit's outcome under a specific treatment.


\section{Stochastic potential outcomes and the stable unit treatment value assumption}

Stochastic models for potential outcomes require a generalization of  SUTVA (the stable unit treatment value assumption; \textcite{rubin1980comment}).  The issue here arises not with the first assumption of SUTVA, the familiar rule of no interference within units, but rather with the lesser-known second assumption, which \textcite{rubin2005causal} expresses as ``there are no hidden versions of treatments; no matter how unit $i$ received treatment 1, the outcome that would be observed would be $Y_i(1)$ and similarly for treatment 0.''  In the Russian roulette example, there are no ``hidden versions'' of treatments in the usual sense of the phrase---you spin the cylinder and fire the shot; it's the same treatment every time---but the potential outcomes are not deterministic.

That is fine.  SUTVA is not a requirement for causal inference; it is just a set of assumptions that are appropriate in some settings but not for others.  Just as we lift the no-interference-within-units assumption when modeling spillover effects, we can lift the second assumption of SUTVA when considering treatments that yield stochastic potential outcomes.  There are many problems in biomedical and social research where outcomes are inherently noisy, so that such stochastic causal models make sense, even for treatments that are applied the same way to each unit.

Stochastic potential outcomes is not a new
idea---\textcite{neyman1935statistical} refers to them as ``technical errors.''  In classical decision theory, they can be thought of as unobserved latent variables, so that any stochastic potential outcome model can be rewritten as a distribution over deterministic values, in the same way that logistic regression can be viewed as an integration over logistically-distributed latent variables.  \textcite{vanderweele2012stochastic} explain how a stochastic potential outcomes model can facilitate a convenient framework for modeling multiple causes.

This modeling choice is reminiscent of the distinction in classical statistical theory between parameters, which are conditioned on when evaluating inferential methods, and latent data or predictions, which are averaged over.  For example, classical statistics distinguishes between unbiased {\em estimation} of parameters $\theta$ (the condition $\mbox{E}(\hat{\theta} (y) | \theta) = \theta$) and unbiased {\em prediction} of latent data $u$ (the condition $\mbox{E}(\hat{u} (y)| \theta) = \mbox{E}(u|\theta)$).  In Bayesian inference, there is no such distinction---all parameters are considered to be a form of latent data and are given probability distributions.  This is consistent with the property of the Neumann-Morgenstern framework (which, confusingly, is often called classical decision theory although it is fully Bayesian) that there is no fundamental distinction between deterministic and stochastic potential outcomes.

Even setting aside the problem of utilities that depend on unrealized potential outcomes, we have found stochastic potential outcomes to be useful as a way of partitioning into unit-specific and unexplained variation \parencite{gelman2021counterexample}.  More formally, one can imagine hypothetical multiple applications of the treatment to the same unit:  in the Russian roulette example, different trials could yield different results on the same person, whereas if the potential outcome depends entirely on person-level characteristics, it would be the same every time.  As discussed in the next section, the partition of variation depends on the extent to which the potential outcomes depend on immutable (if unobserved) unit-level characteristics, which in turn depend on the application.  

\section{Connecting the mathematical and substantive models}

We consider examples to explore the role of uncertainty within and variation between units.
For the Russian roulette example, the problem with the deterministic potential outcome model is that it assigns the ultimate outcome under $R_{p}$ as a property of the individual rather than as a product of inherent randomness in the treatment.  To justify the probabilistic model, one could imagine the spinning to be performed under the control of an external random number generator.

The deterministic and stochastic potential outcome formulations imply the same joint population distribution for $(y^0,y^1)$ but correspond to different models of the individual $(y_i^0,y_i^1)$'s.  When working with classical Neumann-Morgenstern utility theory, these differences have no decision implications because uncertainty about all potential outcomes is integrated out, but when considering asymmetric utilities that depend on unrealized potential outcomes, these differences can affect the decision, as we have seen for the example of comparing $R_{1/6}$ to $R_{1/7}$.


To frame it as a medical problem, suppose that for a certain otherwise-fatal snakebite the current treatment is an antidote that works perfectly except for 1/6 of the population who have a certain genetic condition.  A proposed alternative works perfectly except for 1/7 of the population who have a different condition whose occurrence is statistically independent of the other condition in the population.  Further, suppose that only one antidote can be given to all patients---or, equivalently, that we are treating one patient whose genetic conditions are unknown, and only one antidote can be tried.  The decision of whether to switch to the second treatment is equivalent in expected value to the earlier-considered decision to switch from $R_{1/6}$ to $R_{1/7}$, but it corresponds to variation across units rather than uncertainty within units and thus a different decision recommendation when using the asymmetric utility function that depends on unrealized outcomes.

When mapping this utility function to a decision recommendation, it matters where the conditioning is done.  If you want to use this sort of asymmetric utility function, you cannot simply integrate out the uncertainty of stochastic potential outcomes ahead of time.  The difference between the Russian roulette problem and the snakebite problem is that, in the latter, the potential outcomes are latent variables that are deterministic attributes of the individuals.

How would this apply to real-world outcomes?  We consider two areas:  medicine and business.
\begin{enumerate}
\item In a cleanly specified medical example in which the outcome is binary (survival or death), when would it make sense to prefer to stick with a treatment that leaves more people dead?  There is a vast literature on the trolley problem, and the resolution often turns on issues of agency.  One rationale for preferring the status quo, even if the new treatment increases the expected number of lives saved, is that this sort of ``social engineering'' would reduce people's control over their own lives.  In the Russian roulette example, this reasoning would not apply because the decision to switch from $R_{1/6}$ to $R_{1/7}$ does not harm any identifiable patient.
\item When making a business decision for which a person's negative outcome is not death but rather is to no longer be a customer, there are various reasons to prefer the status quo even if it is stochastically dominated by an alternative treatment.  Making changes (``churn'') has its own costs, and one can easily imagine a scenario in which a loss of 1/7 of a customer base, even if accompanied by a gain of 1/6, would represent a net loss.  This would be the trolley or snakebite scenario.  In the Russian roulette scenario, the decision to switch from $R_{1/6}$ to $R_{1/7}$ would simply result in a reducing the loss by 1/42, which is unambiguously better, as there are no particular customers who could be identified as being lost under one treatment or the other.
\end{enumerate}

To push this further, consider a setting where the potential outcomes are mathematically deterministic but effectively random.  For example, suppose that for some reason a company is about to lose all customers whose Social Security numbers are divisible by 6, and an intervention is proposed by which the company will instead lose all customers whose Social Security numbers are divisible by 7.  Here the potential outcome is deterministic as in the snakebite scenario, but it is also similar to Russian roulette, in that there are no relevant distinguishing characteristics of people in the four principal strata defined by the deterministic outcomes $(y^0,y^1)$.  In this case it could make sense to integrate over this uncertainty, that is, to model the potential outcomes as stochastic.  What this last twist on our example demonstrates is that the appropriate parameterization ultimately depends not just on the asymmetric utility function but also on its underlying motivation.  In real-world settings, we would expect the units in the principal strata to differ in relevant ways, and we expect it would make sense to employ a stochastic potential-outcome model where the distribution for $(y^0,y^1)$ depends on observed and latent characteristics of the experimental units.

We have so far considered the two extreme cases in which variation arises either all from within-unit uncertainty (as with the Russian roulette example) or all from between-unit variation (as with the snakebite example). We will now consider an intermediate example in which both levels of variation are relevant. 
Imagine there are two kinds of patients presenting with headaches. For one type of patient, the headache is a migraine that is not responsive to treatment; for the other the headaches do respond to treatment with some probability of success. The potential outcome is deterministic for the migraine patients but,  depending on how much is known about the mechanism of the treatment, could be stochastic for the non-migraine patients.  This parameterization of potential outcomes would make sense even if individual patients' migraine status is unknown, because the possible action of the treatment depends on characteristics of the individuals.  When comparing to a new treatment that works for some migraine patients but at the cost of reduced efficacy for others, one might want to use an asymmetric utility function that depends on unrealized  potential outcomes. 
Using stochastic potential outcomes to account for the within-unit uncertainty for the non-migraine patients allows us to avoid the inconsistent policy recommendations we encountered in the Russian roulette example, while maintaining the benefits of a ``first do no harm'' policy on the level of across-unit variability. 

\section{Discussion}

There is a large literature on decision rules and utility functions that go beyond classical decision theory.  There could be a general conservatism or disutility of making changes, just because churn has its own costs---or, in other settings, novelty could be considered to having an inherent benefit as a counter to status quo bias.  Preferences can also be based on agency:  people can be more comfortable with risks that they feel they have the power to control and can resist an apparently clear improvement if it represents a perceived loss of power or status.  And meta-decision issues come up, ideas such as satisficing \parencite{simon1958rational}, Type II rationality \parencite{good1971principles}, and fast and frugal heuristics \parencite{gigerenzer1996fast} that recognize the cost of gathering and evaluating the information required to make theoretically optimal decisions.  There is an expression in poker, ``Evaluate the strategy, not the play,'' and one might choose relatively inefficient decision heuristics as part of a global decision-making strategy that is cheaper (less costly to evaluate, exerts a lower cognitive or moral load, etc.).  One could imagine a ``First, do no harm'' principle being morally and socially valuable, to the extent that it could be worth making some objectively bad decisions so as to preserve a larger principle of consistently, in the same way that we like to say we live in ``a government of laws, not of men,'' and so there will be unjust decisions that still need to be enforced in order to not lose that consistency.

The relevance of these ideas to the present paper is that, once we go beyond classical (Neumann-Morgenstern) decision theory and allow the utility of a decision to depend on something other than its outcomes (for example, life or death in the simple medical or Russian roulette example), stochastic potential outcomes can no longer be thought of as equivalent to population averages of discrete potential outcomes.  This is a good thing, in that it allows us to apply the asymmetric utility function of \textcite{benmichael2024policy} and \textcite{christy2024starting} to the trolley or snakebite scenario without having to swallow the uncomfortable conclusion that it is better not to switch from $R_{1/6}$ to $R_{/17}$ in the Russian roulette example.  That is, we can attack those those different problems with the same utility theory, just expressing their potential outcomes differently.

When working with classical (Neumann-Morgenstern) decision theory in which the utility depends only on realized outcomes, there is no fundamental difference between stochastic and deterministic potential-outcome models:  the stochastic model can be transformed into a deterministic model by simply interpreting the joint distribution of $(y^0,y^1)$ as representing a superpopulation distribution from which the deterministic outcomes $(y^0_i,y^1_i)$ are sampled.  It is just a matter of convenience which model is easier to interpret in any given problem (Greenland, 1987).

Once we go beyond classical decision theory and allow utilities that depend on counterfactuals, though, the framing makes a difference.  For decision analysis, going from stochastic to deterministic models for the potential outcomes has the effect of reordering of the steps of expressing uncertainty and evaluating the utility, and outside the Neumann-Morgenstern framework, these steps do not commute.

The Russian roulette problem is a simple example demonstrating how the relative utility of different decisions---even the ranking of which decision is preferred---can depend on how the potential outcomes are parameterized.  Although this might seem awkward, it is a mathematical consequence of the use of a utility function that depends on unrealized outcomes.

To the extent that such decision rules are desirable, it behooves us to think of this dependence on parameterization as a feature rather than a bug.  A principle such as ``first do no harm'' depends on what is considered ``harm,'' which in turn depends on what aspects of a problem are considered to be potentially under control. By using potential outcomes that have stochastic components, we are making such decisions explicit.

\section*{Acknowledgement}
We thank Amanda Kowalski, the editor, and two reviewers for helpful discussions and the U.S. Office of Naval Research for partial support.

\printbibliography

\end{document}